\documentclass[aps,prx,superscriptaddress,twocolumn,showpacs]{revtex4-1}

\usepackage{amsmath, amsthm, amssymb}
\usepackage{graphicx}
\usepackage{color}
\usepackage{times}
\usepackage{dsfont}
\usepackage{epstopdf}

\usepackage{natbib}
\usepackage{hyperref}
\hypersetup{
        colorlinks=true,
}

\newcommand{\id}{\mathds{1}}

\newcommand{\eqnref}[1]{(\ref{#1})}

\newcommand{\ket}[1]{\mbox{$| #1 \rangle$}}

\newcommand{\subhead}[1]{\vspace{0.15in} \noindent \textbf{#1} \\}

\begin{document}

\title{Chiral spin liquid and emergent anyons in a Kagome lattice Mott insulator}

\author{B. Bauer}
\affiliation{Station Q, Microsoft Research, Santa Barbara, CA 93106-6105, USA}

\author{L. Cincio}
\affiliation{Perimeter Institute for Theoretical Physics, Waterloo, Ontario, N2L 2Y5, Canada}

\author{B. P. Keller}
\affiliation{Physics Department, University of California,  Santa Barbara, CA 93106, USA}

\author{M. Dolfi}
\affiliation{Theoretische Physik, ETH Zurich, 8093 Zurich, Switzerland}

\author{G. Vidal}
\affiliation{Perimeter Institute for Theoretical Physics, Waterloo, Ontario, N2L 2Y5, Canada}

\author{S. Trebst}
\affiliation{Institute for Theoretical Physics, University of Cologne, 50937 Cologne, Germany}

\author{A. W. W. Ludwig}
\affiliation{Physics Department, University of California,  Santa Barbara, CA 93106, USA}

\begin{abstract}
Topological phases in frustrated quantum spin systems have fascinated researchers for decades.
One of the earliest proposals for such a phase was the chiral spin liquid put forward by Kalmeyer and
Laughlin in 1987 as the bosonic analogue of the fractional quantum Hall effect. Elusive for many years,
recent times have finally seen a number of models that realize this phase. However, these
models are somewhat artificial and unlikely to be found in realistic materials. Here, we take an
important step towards the goal of finding a chiral spin liquid in nature by examining a physically
motivated model for a Mott insulator on the Kagome lattice with broken time-reversal symmetry.
We first provide a theoretical justification for the emergent chiral spin liquid phase in terms of a
network model perspective. We then present an unambiguous numerical identification and
characterization of the universal topological properties of the phase, including ground state degeneracy, edge
physics, and anyonic bulk excitations,
by using a variety of powerful numerical probes, including the entanglement spectrum and modular
transformations.

\end{abstract}

\maketitle

\section{Introduction}

The low-energy properties of frustrated quantum spin systems -- loosely speaking,
systems of interacting spins in which the local energetic constraints cannot
all be simultaneously satisfied -- have fascinated researchers for
many decades. These systems arise in the description of the spin degrees
of freedom of Mott insulators, i.e. insulating states where the fluctuations of
the charge degrees of freedom have been suppressed by interactions
while the spin degrees of freedom remain free to form non-trivial quantum
phases. Such states are found in many materials, but can also be
artificially made in the lab using cold atomic gases~\cite{greiner2002,jordens2008}.
In most situations, the spins collectively order into some pattern
that can be described through a local order parameter. A more exciting possibility,
coined {\it spin liquid phase}~\cite{balents2010}, is that the spins do not order into
such a local pattern; instead, a more exotic state governed by strong quantum
fluctuations emerges.

In a famous paper in 1987, Kalmeyer and Laughlin~\cite{kalmeyer1987}
hypothesized a scenario where a chiral topological
spin liquid (CSL) is formed. In this elusive phase of matter, the spins form
a collective state that can only be described in terms of emergent, non-local
topological properties. So far, this behavior has experimentally only been observed in fractional
Quantum Hall systems~\cite{tsui1982,laughlin1983}.
Such topologically ordered liquids~\cite{wen1990-1} are characterized through a
number of universal properties ranging from topologically protected gapless
edge states~\cite{halperin1982,wen1990} and a ground state degeneracies
that depend on the topology of the sample~\cite{wen1990-1} to exotic
excitations that carry fractional charge and satisfy neither fermionic nor
bosonic exchange statistics~\cite{moore1991}. These \emph{anyonic particles}~\cite{wilczek1982}
can serve as key ingredients in topological quantum computers~\cite{nayak2008},
making them relevant also for technological applications.

In the specific case of the chiral spin liquid as proposed by Kalmeyer and
Laughlin, the universal properties of the
ground state are captured by the bosonic $\nu=1/2$ Laughlin
state~\cite{laughlin1981}. Probably the most striking property of this state
are its semionic bulk excitations: When exchanging two such semions,
the wave function describing the collective state of the system acquires
a complex phase $i$, in stark contrast
to conventional bosons or fermions, where the factor is 1 and $-1$,
respectively. Equally striking is the emergence of a topologically
protected chiral edge state with a universal spectrum at the boundary
of the sample. This leads to unidirectional transport along the
boundary of the sample, while the bulk remains insulating. The 
correspondence between edge and bulk physics has
been used as a powerful experimental probe into the physics of
fractional quantum Hall systems. Finally, the
ground state degeneracy of this state depends on the topology of the
underlying manifold; for example, when placed on a torus, two ground
states are found.

\begin{figure}
  \includegraphics[width=1.5in]{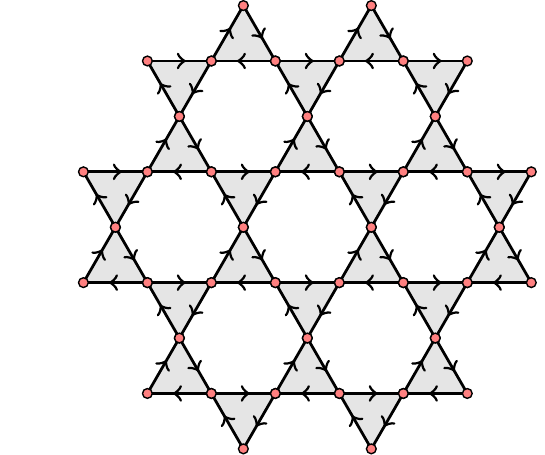}
  \includegraphics[width=1.5in]{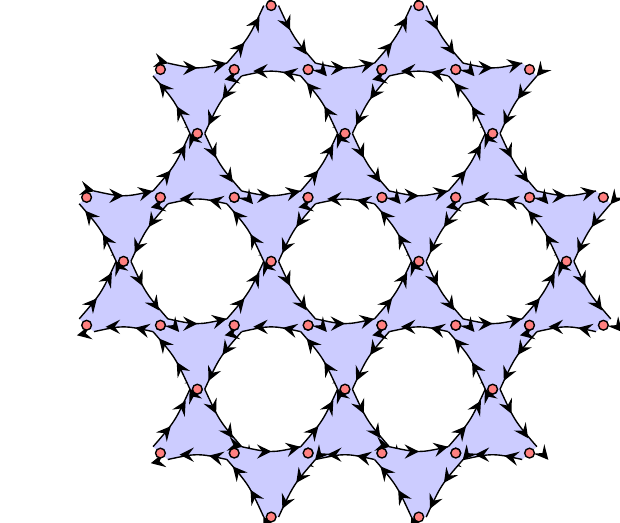}
  \caption{
  {\it Left panel:} Kagome lattice considered in this manuscript, where grey shading
  indicates the elementary triangles. Arrows on the bonds indicate the direction
  induced by the magnetic flux $\Phi$ enclosed in each triangle. 
  {\it Right panel:} Visualization of the network-model perspective on the chiral spin liquid phase arising from Hamiltonian~\eqnref{eqn:ham}. Consistent with a chiral topological phase, a collective edge state encircles the whole systems. In this particular model, additional closed edges encircle each hexagon.
  \label{fig:kagome}
  }
\end{figure}

Over the last decades, much research has been devoted to finding
realistic spin Hamiltonians that have such a chiral topological phase
as their ground state, but to this date, the only known examples are
Hamiltonians that are unlikely to be relevant for any material~\cite{yao2007,schroter2007,tang2011,sun2011,neupert2011,nielsen2013}.
Here, we study a simple spin model on the Kagome lattice (left panel of Fig.~\ref{fig:kagome}) that is derived from
the Hubbard model, which is the minimal relevant
model for itinerant interacting electrons,
in the presence of time-reversal symmetry breaking.
The Hubbard Hamiltonian reads
\begin{equation} \label{eqn:hubb} \begin{split}
H =& - \sum_{\langle i,j \rangle, \sigma} (t_{ij} c_{i \sigma}^\dagger c_{j \sigma} + t_{ij}^* c_{j \sigma}^\dagger c_{i \sigma}) \\
& + \frac{h_z}{2} \sum_i (n_{i \uparrow} - n_{i \downarrow}) + U \sum_{i} n_{i \uparrow} n_{i \downarrow}.
\end{split} \end{equation}
Here, a magnetic field induces both a Zeeman term $h_z$ as well as a flux $\Phi$ through
each elementary triangle of the lattice, such that for $i,j,k$ clockwise around a triangle
we have $t_{ij} t_{jk} t_{ki} = t^3 \exp(i \Phi)$, as indicated by the arrows in the left panel of
Fig.~\ref{fig:kagome}.
When $\Phi \neq 0$ or $h_z \neq 0$, time-reversal symmetry is broken.
When considered at half-filling, $\langle n \rangle = 1$, and in the limit
of large repulsive interaction strength $U$, a Mott insulating state is formed and an effective
spin model can be derived from perturbation theory in $t/U$~\cite{motrunich2006}.

Here, we will demonstrate conclusively that in a very wide parameter regime where
a large enough magnetic flux $\Phi$ breaks time-reversal symmetry,
the ground state of the effective spin model -- and hence also the Hubbard model in an
appropriate parameter regime -- is a chiral topological spin liquid with
emergent anyonic excitations.

\section{Model} \label{sec:model}

Starting from from the Hubbard model of Eqn.~\eqnref{eqn:hubb}, a $t/U$ expansion
at half filling gives the following spin Hamiltonian~\cite{motrunich2006}:
\begin{equation} \label{eqn:ham} \begin{split}
H =&~ J_\text{HB} \sum_{\langle i,j \rangle} \vec{S}_i \cdot \vec{S}_j + h_z \sum_i S^z_i \\
& +J_\chi \sum_{i,j,k \in \bigtriangleup} \vec{S}_i \cdot (\vec{S}_j \times \vec{S}_k) + \ldots,
\end{split} \end{equation}
where for the three-spin term, the $i,j,k$ are ordered clockwise around the elementary triangles of the
Kagome lattice. The term $\chi_{ijk} = \vec{S}_i \cdot (\vec{S}_j \times \vec{S}_k)$, referred to
as the scalar spin chirality~\cite{wen1989,baskaran1989},
breaks time-reversal symmetry and parity, but preserves SU(2) symmetry.
To lowest order, the coupling parameters depend on the parameters of
the Hubbard model as $J_\text{HB} \sim t^2/U$ and $J_\chi \sim \Phi t^3/U^2$,
ignoring further subleading terms. We choose
to parametrize the model using $J_\text{HB} = J \cos \theta$ and $J_\chi = J \sin \theta$
and set $J=1$.

In the absence of time-reversal symmetry breaking ($\theta=0$ and $h_z=0$),
this is the Kagome lattice nearest-neighbor Heisenberg antiferromagnet, which has become
a paradigmatic model for frustrated magnetism~\cite{elser1989,marston1991,sachdev1992} with possible
spin liquid ground state and relevance to the description of materials such as Herbertsmithite and
Volborthite~\cite{lee2007,han2012}.
Recent numerical work~\cite{yan2011,jiang2012-1,depenbrock2012} has indicated 
that this model may realize a time-reversal symmetric $\mathbb{Z}_2$ topological spin liquid,
while other numerical results give evidence for a gapless spin liquid phase~\cite{iqbal2011,clark2012}.

Here, we explore the ground state phase diagram of~\eqnref{eqn:ham} away from
the time-reversal invariant Heisenberg point $\theta=0$. In particular, we find an
extended chiral spin liquid phase around the point $\theta=\pi/2$ and $h_z=0$,
where the Hamiltonian reduces to the three-spin term,
\begin{align} \label{eqn:csl-ham}
H_{\text{CSL}} &= \sum_{i,j,k \in \bigtriangleup} \chi_{ijk}.
\end{align}
Our numerical results indicate that the CSL is in fact stable almost up to the
Heisenberg point, namely for all $\theta \geq 0.05 \pi$.
We also establish an extended range of stability against other perturbations, including
(i) the Zeeman field,
(ii) an easy-axis spin anisotropy in the Heisenberg term,
(iii) a next-nearest neighbor Heisenberg term, and
(iv) the Dzyaloshinsky-Moriya (DM) interaction induced by Rashba-type spin-orbit coupling for the fermions.
While the aim of this paper is not to examine the nature of the
transitions out of the chiral spin liquid phase, e.g. the expected phase transition
to the time-reversal symmetric spin liquid of the Heisenberg antiferromagnet,
these questions should be addressed in future work.

In the following, we will use two complementary routes to show that the ground
state of Eq.~\eqnref{eqn:csl-ham} is indeed a chiral spin liquid.
First, we argue for this from a powerful perspective rooted in the physics of network models of
edge states akin to the Chalker-Coddington network model for the integer
quantum Hall transition~\cite{chalker1988}. We will then turn to powerful numerical
tools to unambiguously identify the universal properties of the chiral spin liquid.

\begin{figure}
  \begin{tabular}{c}
  \includegraphics[width=3in]{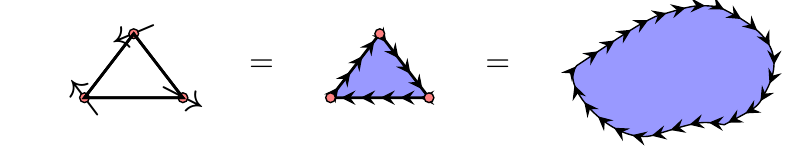} \\ \hline
  \vspace{0.1in} \includegraphics[width=2.5in]{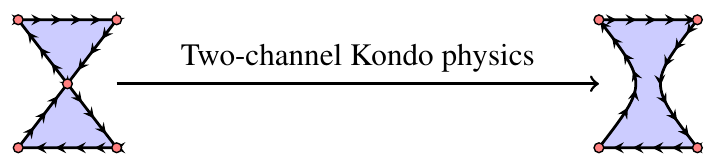}  
  \end{tabular}
  \caption{(color online)
  {\it Top:} Sketch of a puddle of topological phase replacing each triangle of three spins.
  {\it Bottom:} Behavior of two corner-sharing triangular puddles of the topological phase. \label{fig:puddle} }
\end{figure}

The key step of our first argument is to view each triangle of spins
as the seed of a chiral topological phase, a puddle encircled by
an edge state, as illustrated in the top panel of Fig.~\ref{fig:puddle}. The natural
candidate for the phase filling the puddle is the bosonic $\nu=1/2$
Laughlin state~\cite{halperin1983}\nocite{moore1991}, which is the simplest
bosonic quantum Hall state known to possess the SU(2) symmetry required by our construction.
It is also the state envisioned by Kalmeyer and Laughlin~\cite{kalmeyer1987}.
Forming a lattice out of the elementary triangles, we should then consider
a situation with many individual puddles of this topological phase. To
see what collective state is formed, we have to understand how two
corner-sharing triangles of the Kagome lattice are joined.
This situation of edges meeting at the corner spin shared by two triangles
is an incarnation of two-channel Kondo physics~\cite{AffleckLudwig1991,MaldacenaLudwig1997},
for which it is well-known that the edges will {\it heal}~\cite{eggert1992,kane1992}
if the coupling
to the center spin is symmetric, as illustrated in the lower panel
of Fig.~\ref{fig:puddle} and discussed in more detail in the appendix. Thus, the corner spin has
merged the two triangles to form a larger puddle encircled by
a single edge state, i.e. to form a larger region of the topological phase.
We can repeat the above step (Fig.~\ref{fig:puddle}) for all pairs of
corner-sharing triangles of the Kagome lattice. The system then forms
one macroscopic, extended region of a \emph{single} topological
phase with one edge state encircling its outer boundary, as illustrated
in the right panel of Fig.~\ref{fig:kagome}, and with closed loops encircling the interior
hexagons of the Kagome lattice. We thus obtain a direct realization of the
Kalmeyer-Laughlin state for a chiral topological spin liquid phase.

\section{Numerical identification of the CSL} \label{sec:numerics}

We now turn to a numerical identification of the CSL 
at the chiral point $\theta=\pi/2$ of Hamiltonian~\eqnref{eqn:ham} and
in its vicinity by studying the three
hallmark properties of a chiral topological phase: the presence of
(i) a gapped spectrum with a topological degeneracy on the torus,
(ii) a gapless edge state with a universal spectrum of low-energy excitations,
and (iii) anyonic bulk excitations.
On a technical level, we resort to exact diagonalization and DMRG calculations
to extract energy spectra, entanglement spectra, and modular matrices for
various system configurations.
To label their diameter and boundary condition, we will
use the notation introduced in Ref.~\onlinecite{yan2011}; see also
the Methods section.

\begin{figure}
  \includegraphics[width=\columnwidth]{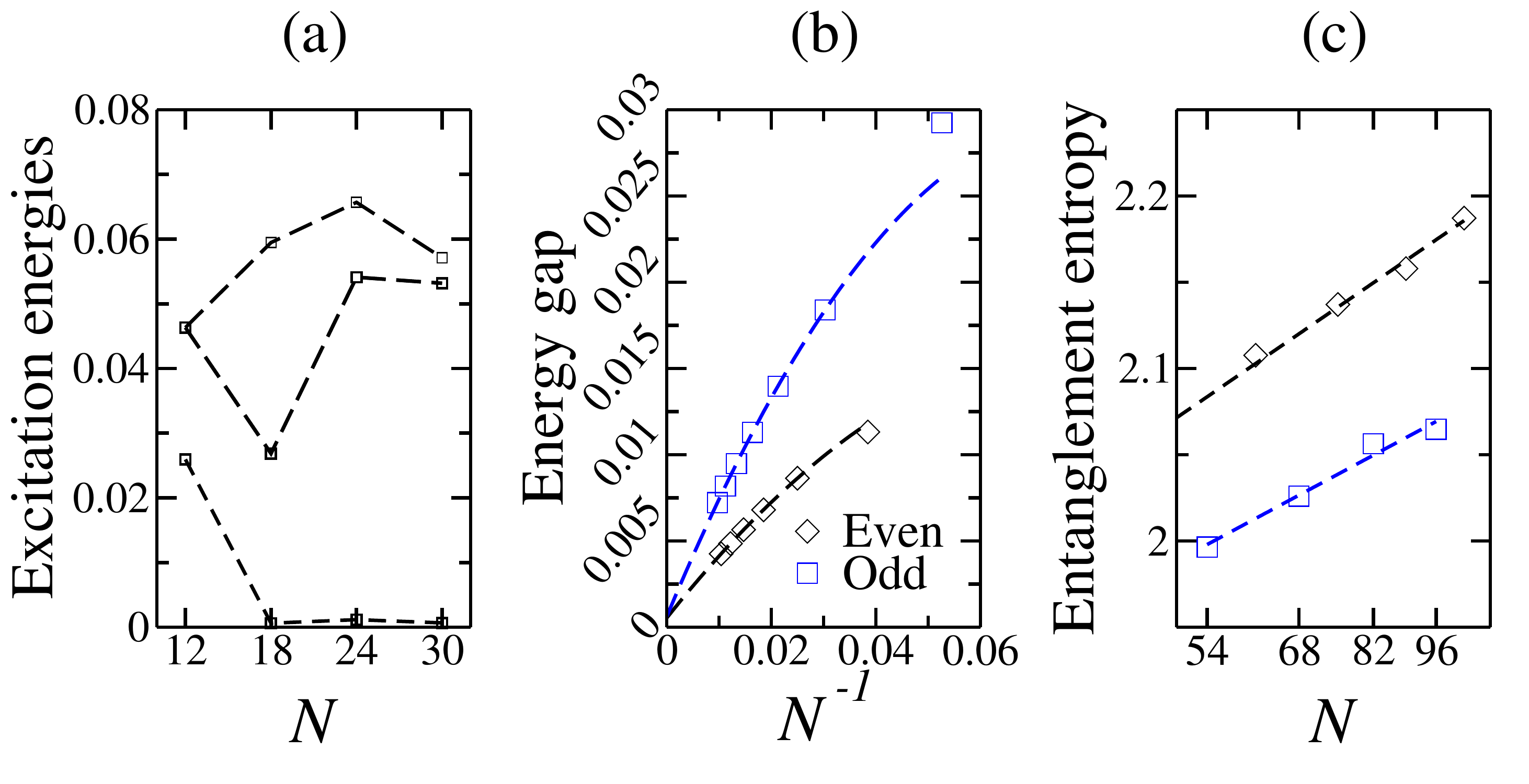}
  \caption{
  {\it (a)} Exact diagonalization excitation energies on small tori of type XT4-0 up to length 5. 
  {\it (b)} Energy gap on a thin, long strip of width 4 sites; dashed lines indicate the extrapolation to $N \rightarrow \infty$. The two different branches denote systems with an even (black) and odd (blue) number of unit cells.
  {\it (c)} Entanglement entropy at the center of the same system as (b) on a semi-logarithmic scale for even (black) and odd (blue) number of unit cells; dashed lines indicate a fit. Both fits are consistent with $c=1$.
  In all panels, $N$ is the total number of lattice sites.
  \label{fig:spectrum} }
\end{figure}

We first demonstrate that the system has a finite gap in the thermodynamic limit.
To this end, we consider a sequence of XT4-0 tori of length up to 5 unit cells, shown in the left panel
of Fig.~\ref{fig:spectrum}.
For systems with $N \geq 18$ sites, there clearly is a low-lying excitation, which can be attributed
to a two-fold near-degeneracy of the ground state. All excitations above these near-degenerate ground
states are separated by a spectral gap of roughly $\Delta \approx 0.05$.
Further consistent evidence for the gap can be obtained from exact diagonalization of a 36-site
XT6-3 cluster and on XT4-2 clusters of size up to 30 sites (not shown).
As a further consistency check, we can extract the gap for long, thin cylinders using DMRG.
Performing this for cylinders of type XC4-0 with up to 100 sites,
we confirm that the triplet gap does not depend significantly on the length
of the system, ruling out the presence of gapless modes propagating along the cylinder.

We conclude from this that the gap remains finite in the thermodynamic limit. The qualitative
agreement between the spectral gap extracted for different system sizes and boundary conditions
is also a strong indication
that the correlation length of the system is short compared to the system sizes we are able to
study numerically. To further support this, we have calculated the spin-spin and dimer-dimer
correlation functions as well as an upper bound on the asymptotic correlation
length on infinite cylinders. All of these indicate a correlation length on the order
of {\it one} unit cell. Taken together, this gives strong evidence that the properties observed on
small tori and quasi-one-dimensional systems are representative of the
two-dimensional phase in the thermodynamic limit.

The observed two-fold near-degeneracy is consistent with what is
expected for the CSL, namely a
$2^g$-fold ground state degeneracy on a manifold of genus $g$, which
will be split by an amount that is exponentially small in the system size.
We also find two states $\ket{\Psi_a}$ with very similar energy densities for infinite
cylinders of type XC8-4 and XC12-6. As explained in the
Methods section, the two states can be identified by
their well-defined topological flux $a$ through the cylinder,
which for the $\nu=1/2$ Laughlin state can be the identity
($a = \id$) or a semion ($a=\mathrm{s}$).

\subhead{Edge physics}
Placing a chiral topological phase on a cylinder or disk, a gapless chiral edge state emerges with
a universal spectrum governed by a conformal field theory~\cite{halperin1982,wen1990}.
To observe this, we consider the spectral gap of the system
when placed on a thin, long strip with a fixed width of 4 sites.
In stark contrast with the case of a thin long cylinder, the spin gap
vanishes as $a/L+b/L^2$, where $a$ and $b$ are
parameters of the fit (Fig.~\ref{fig:spectrum}(b)).
This is a hallmark signature of a gapless edge mode. We can further
pinpoint the universality class of the edge theory by extracting its central charge $c$
from the entanglement entropy. As shown in the Fig.~\ref{fig:spectrum}(c), we find good agreement
with a value of $c=1$, which is precisely that expected for the chiral SU(2)$_1$ Wess-Zumino-Witten
conformal field theory describing the edge of a $\nu=1/2$ Laughlin state.

\begin{figure}
  \includegraphics[width=\columnwidth]{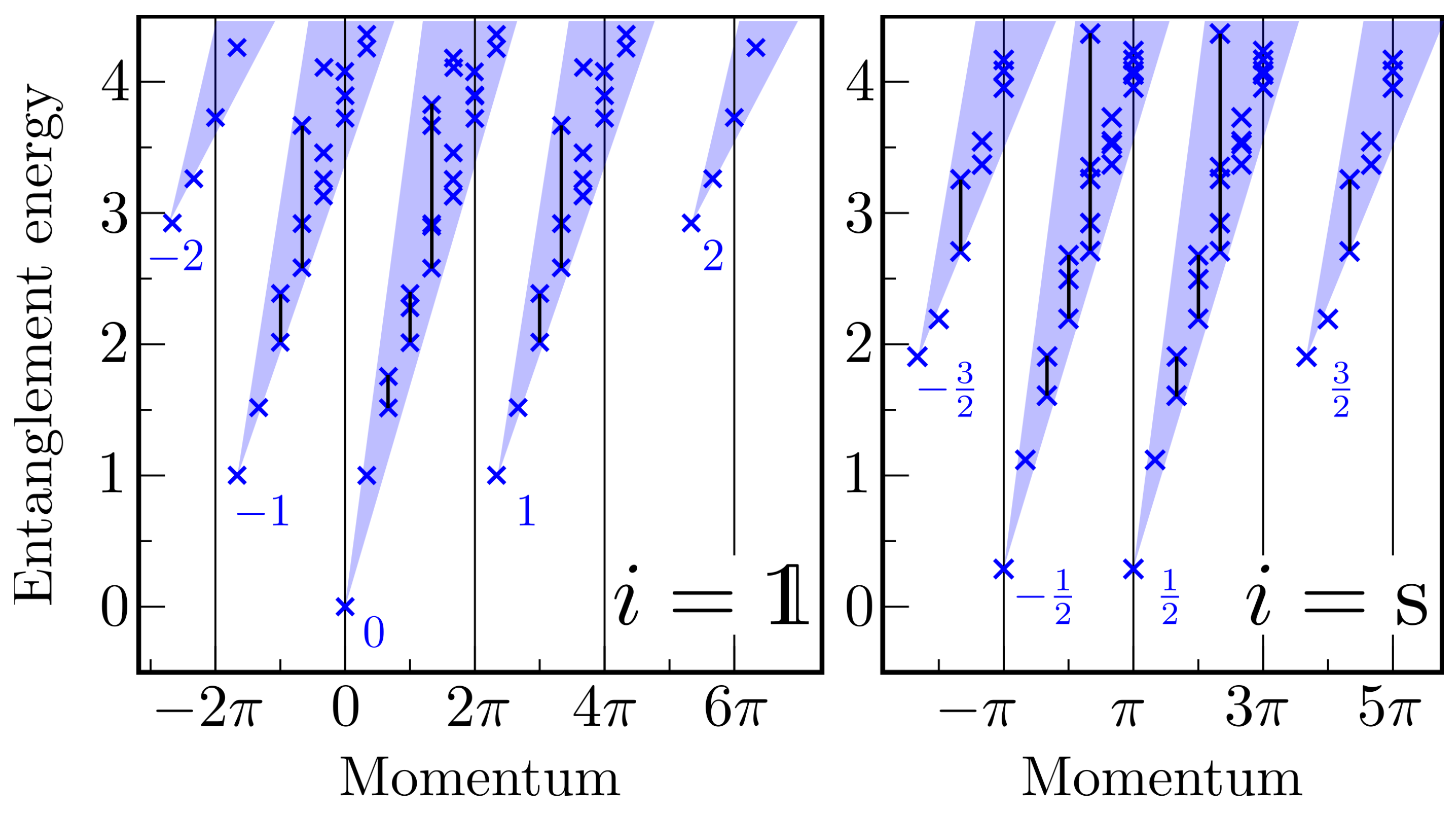}
  \caption{Entanglement spectrum of the reduced density matrix $\rho_a$ for one half of an infinite cylinder obtained for both ground states $\ket{\Psi_\id}$ (left) and $\ket{\Psi_\mathrm{s}}$ (right panel). The entanglement energies shown on the vertical axis, up to the global shift and rescaling, are given by $E_{a,\sigma} = -\log (p_{a,\sigma})$, where $p_{a,\sigma}$ are the eigenvalues of $\rho_a$. The horizontal axis shows the momentum in the transverse direction of the corresponding eigenvectors of $\rho_a$. Each tower is identified by its $S^z$ quantum number as indicated by the blue label; we have offset the momentum of different towers by $2\pi$ to improve clarity. The cylinder used here is XC12-6.
 }
 \label{fig:cft}
\end{figure}

As an even more refined probe, we use the entanglement spectrum, which
reflects the same universal properties as the physical edge spectrum~\cite{li2008,qi2012,chandran2011,dubail2012,swingle2012}.
For each of the two ground states $\ket{\Psi_a}$ obtained for an infinite cylinder,
the entanglement spectrum, see Fig.~\ref{fig:cft}, is consistent with
the corresponding sector of the
chiral SU$(2)_1$ Wess-Zumino-Witten conformal field theory:
The entanglement spectrum of $\ket{\Psi_\id}$ displays precisely the sequence
of degeneracies of the tower of Kac-Moody descendants
of the identity primary field (1-1-2-3-5-...). These are reproduced by counting
the number of low-lying close-by states in each tower grouped by momentum and spin quantum numbers.
Similarly, the entanglement spectrum of $\ket{\Psi_\mathrm{s}}$ displays the degeneracies of
the spin-1/2 primary field and its descendants (also 1-1-2-3-5-...).
We note that in the identity sector, all towers carry integer representations of the spin quantum number, whereas in the semion
sector they carry half-integer representations. In both ground states, the levels can be grouped into
SU(2) multipletts.

\subhead{Emergent anyons}
The bulk of the chiral spin liquid phase has anyonic excitations, referred to as semions.
The topological properties of these quasiparticles can be characterized through their modular
$T$ and $S$ matrices~\cite{nayak2008}. The $T$ matrix contains the
central charge $c$ and the self-statistics of the anyonic particles, i.e. the phase
that is obtained when two particles of the same kind are exchanged.
The $S$ matrix contains the mutual statistics of the anyonic quasiparticles, their
quantum dimensions (counting the internal degrees of freedom
of each particle), and the total quantum dimension of the phase. More detailed
definitions of these quantities are given in the Methods summary.

For a fixed number of quasiparticles, only a finite number of
possible $S$ and $T$ matrices exist~\cite{rowell2009,bruillard2013}. For two types of
quasiparticles (as in the case of the $\nu=1/2$ bosonic Laughlin state) only two choices are possible~\cite{rowell2009}.
Therefore, by numerically calculating the $S$ and $T$ matrices and
comparing them against the two possibilities, we have fully identified
the universal properties of the topological phase.
For the $\nu=1/2$ Laughlin state, the modular matrices are
\begin{align} \label{TS}
T&=e^{-i \frac{2\pi}{24}} \left[\begin{matrix} 1 & 0 \\ 0 & i \end{matrix} \right]
&S&=\frac{1}{\sqrt{2}}\left[\begin{matrix} 1 & 1 \\ 1 & -1 \end{matrix}\right].
\end{align}

For an XT8-4 torus of $48$ sites at $\theta=0.05 \pi$, where the
finite-size corrections to this quantity are minimal, we obtain
\begin{eqnarray}
T &=& e^{-i \frac{2\pi}{24} 0.988} \left[
\begin{matrix}
1 & 0 \\
0 & i \cdot e^{-i0.0021 \pi}
\end{matrix}
\right] \ , \label{numT} \label{numS} \\
S &=& \frac{1}{\sqrt{2}} \left[
\begin{matrix}
0.996 & 0.995 \\
0.996 & -0.994 e^{-i 0.0019 \pi}
\end{matrix} \right] \ . \nonumber
\end{eqnarray}
This is in very good agreement with the $T$ and $S$ matrices for the $\nu=1/2$ Laughlin state
given in Eqn.~\eqnref{TS} and provides the strongest confirmation of the nature of the
bulk topological phase.
The correct normalization of the first row or column of the $S$ matrix
indicates that we have indeed obtained a full set of ground states.
We can also read off the total quantum dimension $\mathcal{D}=1/S_{\id \id}=\sqrt{2}/0.996$ of the phase,
which determines the topological entanglement entropy~\cite{kitaev2006-tee,levin2006}
that has been widely used to identify topological phases.
Furthermore, the central charge $c=0.988$ is in excellent agreement with the prediction
and the value extracted from the edge above.

\begin{figure}
  \includegraphics[width=\columnwidth]{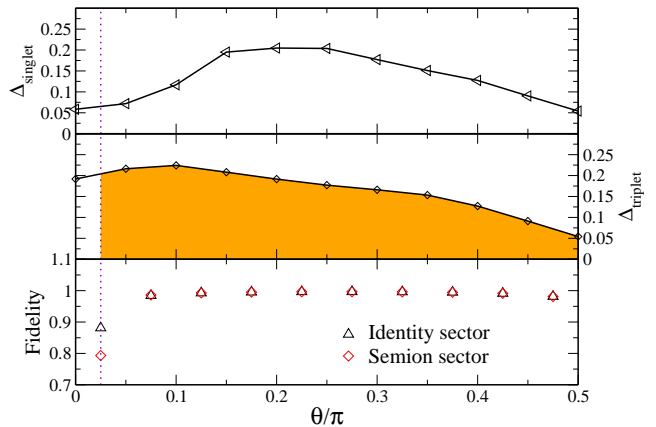}
  \caption{
  Singlet and triplet gaps as a function of $\theta$ for an infinite XC8-4 cylinder.
  The triplet gap is a lower bound on the critical magnetic field $h_c$; hence, the shaded region in
  the middle panel indicates the minimal stability of the phase in the $\theta$-$h_z$ phase diagram.
  \label{fig:pd} }
\end{figure}

\subhead{Stability of the chiral spin liquid}
To establish the region in which the phase persists as $\theta$ is tuned in the range $\theta \in [0,\pi/2]$,
we first consider the fidelity~\cite{zanardi2006} $F(\theta) = \langle \Psi_a(\theta-\epsilon) | \Psi_a(\theta+\epsilon) \rangle$
shown in the bottom panel of Fig.~\ref{fig:pd} (for the precise definition of this quantity for infinite
systems, see the appendix). The fidelity remains near unity as $\theta$ is tuned away from
the chiral point $\theta=\pi/2$, until it suddenly drops for $\theta < 0.05 \pi$, indicating a transition.
Further support for the extended stability of the CSL is found in various other characteristics,
including the spectral gap, the modular matrices and the entanglement spectrum.
This is remarkable as it indicates that tuning away from the Heisenberg model ($\theta=0$)
with a small critical chiral coupling of $(J_\chi/J_\text{HB})_\text{crit} \leq \tan(0.05 \pi) \approx 0.16$
is sufficient to drive the system into the chiral phase. 

In experimental scenarios, a Zeeman magnetic field $h_z$ is likely to be generated along with
the orbital magnetic field that induces the three-spin chiral term. The relative strength of the orbital
magnetic field and the Zeeman field is determined by the $g$-factor and the ratio $t/U$.
The energy gap in the triplet sector gives a lower bound on the critical field strength $h_c$ up 
to which the CSL phase is stable. The values for the triplet gap shown in the middle panel of Fig.~\ref{fig:pd}
remain large all the way from the fully chiral point ($\theta=\pi/2$) to the transition out of the
CSL towards the Heisenberg point ($\theta=0$).

In the top panel of Fig.~\ref{fig:pd}, we also show the singlet gap, i.e. the gap to the lowest excitation
in the $S_z=0$ sector. As opposed to the triplet gap, which appears to remain large across the
transition, the singlet gap decreases as the transition out of the CSL is approached. Note that the gaps may
be rounded off at the transition by effects due to either the finite diameter or the finite bond dimension
of the matrix-product state ansatz.
We point out, however, that a closing only of the singlet and not the triplet
gap is consistent with the scenario for the transition from the chiral spin liquid into a
doubled semion phase (twisted $\mathbb{Z}_2$ topological phase) studied in Ref.~\onlinecite{barkeshli2013}.

\section{Outlook}

We have taken an important step towards finding realistic models for a chiral spin
liquid in a frustrated spin system. 
We believe that this will nucleate new research efforts both by theorists and
experimentalists. From a theoretical point of view, studying the transition from
the chiral spin liquid to the putative time-reversal symmetric spin liquid in the
Heisenberg model will provide the unique opportunity to study a topological
phase transition in a realistic model, and may provide invaluable insights into
the physics of frustrated spins on the Kagome lattice. For experimentalists, our
work will provide a guide in searching for realizations of bosonic fractional
Quantum Hall physics in the lab, be it in materials that have Kagome lattice
structure and form Mott insulators, or by engineering such systems in cold atomic gases.

 \vspace{0.15in}
{\small During completion of this work, we became aware of related work in Refs.~\onlinecite{he2013,gong2013}.}

\acknowledgements
The DMRG code was developed with support from the Swiss Platform for High-Performance
and High-Productivity Computing (HP2C) and based on the ALPS libraries~\cite{bauer2011-alps}.
S.T. was supported, in part, by  SFB TR 12 of the DFG. A.W.W.L was supported, in part, by
NSF DMR-1309667. We acknowledge illuminating discussions with participants of the KITP
workshop \emph{Frustrated Magnetism and Quantum Spin Liquids} (Fall 2012) as well as
M. Barkeshli and P. Bonderson. This research was supported in part by Perimeter Institute for
Theoretical Physics. Research at Perimeter Institute is supported by the Government of Canada
through Industry Canada and by the Province of Ontario through the Ministry of Research and Innovation.

\appendix

\section{Method summary}

\begin{figure}
  \includegraphics[width=2in]{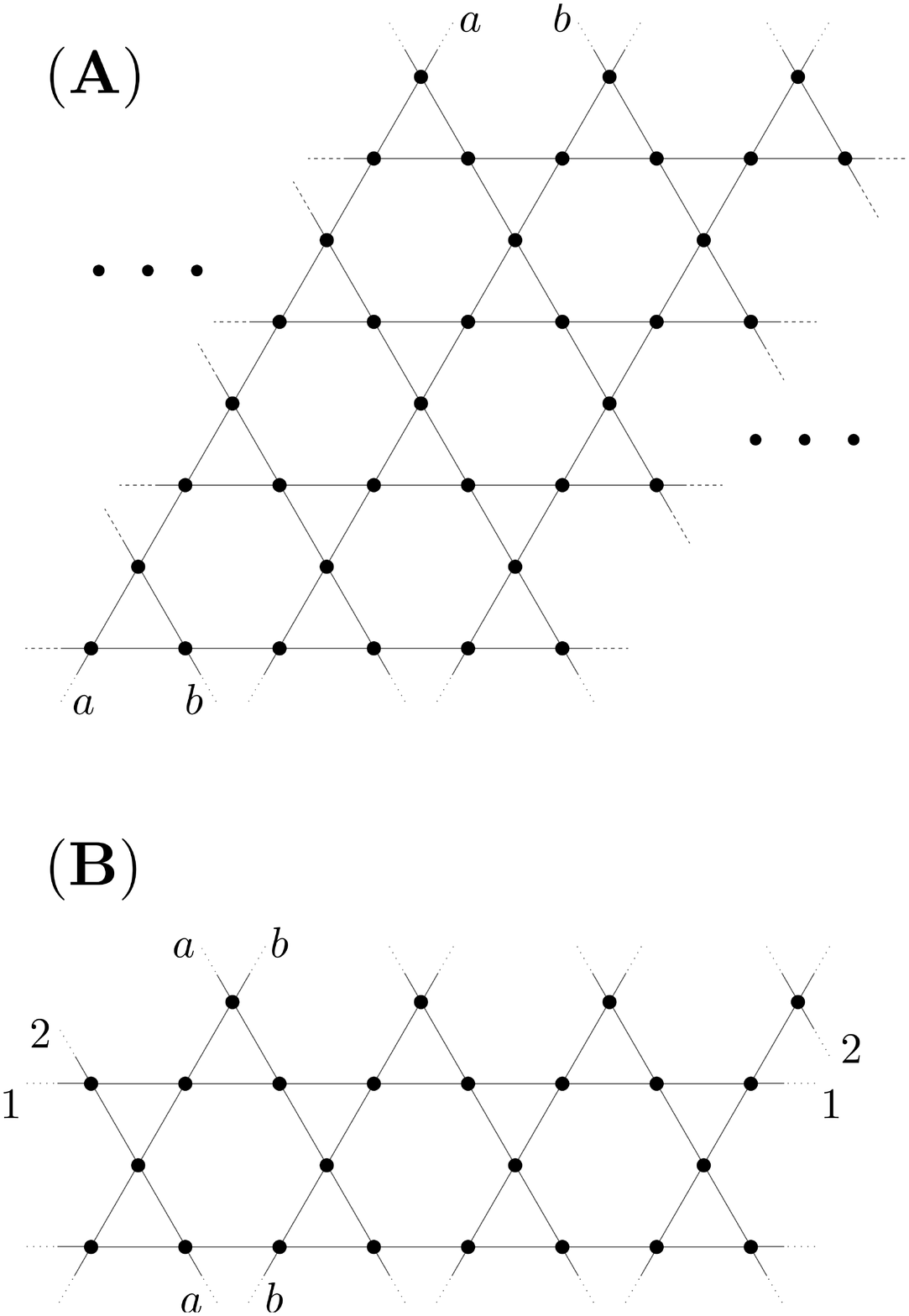}
  \caption{
  {\it Panel (A)}: Section of an XC8-4 cylinder. Sites labeled $a$, $b$ are identified.
  {\it Panel (B)}: XT4-0 torus with 24 sites.
  \label{fig:systems} }
\end{figure}

We label the system sizes according to the scheme introduced in Ref.~\onlinecite{yan2011}. For cylinders,
the labels are XC$m$-$n$; here, X indicates the orientation of the cylinder, $m$ is the diameter of the cylinder
in measured sites, and $n$ is the shift of the boundary condition when wrapping the place to a cylinder. Note that
cylinders of type XC$(2n)$-$n$ are spanned by the basis vectors of the triangular lattice. For tori, we use the
labels XT$m$-$n$, where $m$ and $n$ have the same meaning as above. Two examples are shown in
Fig.~\ref{fig:systems}.

We numerically study finite systems using exact diagonalization as well as the density-matrix
renormalization group (DMRG) method~\cite{white1992,white1992-1}.
While our exact diagonalization is limited to 36 sites, DMRG allows us to study systems in a limit
of long, thin cylinders and strips. Up to $M=4000$ states are used in the finite-size DMRG calculations.
Finite-size DMRG calculations allow us to extract the spin gap as well as the entanglement entropies
for continguous blocks of sites, which we use to extract the central charge of the system by fitting
to the well-known formulas of Ref.~\onlinecite{calabrese2004}.

Furthermore, we use infinite-size DMRG with a translationally invariant matrix product state ansatz of up to $M=4096$
states. This allows us to obtain a complete set of ground states with well-defined anyonic flux,
as first proposed in Ref.~\onlinecite{cincio2012}. Here, the use of an infinite cylinder
(as opposed to a finite cylinder with boundaries, where there is no ground state degeneracy) is key
to obtain one ground state for each anyonic charge, analogous to the torus.
In this basis, we can extract the entanglement spectrum as well as the $T$ and $S$ matrices.
Ref.~\onlinecite{cincio2012} details the non-trivial step of resolving the entanglement spectrum
by transverse momentum. We here only describe how to obtain the $T$ and $S$ matrices and
defer the discussion of how we extract the fidelity and gaps to the appendix.

The $T$ and $S$ matrices are defined through a world-line diagram for anyonic particles as:
\begin{eqnarray*}
&\includegraphics{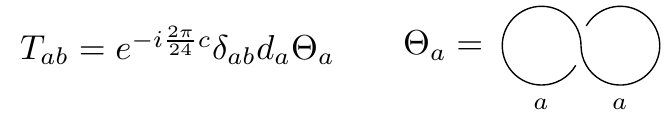} \\
&\includegraphics{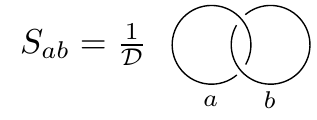}
\end{eqnarray*}
Here, $\mathcal{D}$ is the total quantum dimension of the phase, and $d_a$ is the quantum
dimension of a quasi-particle of topological charge $a$. These matrices can be related to
generators of the modular group on the torus, and fulfill the conditions $(ST)^3=S^2$ and $S^4 = \id$.
The $T$-matrix is proportional to the twist factors $\Theta_a$, which contain the self-statistics
of the excitations, i.e. the phase that is obtained when two particles of the same kind
are exchanged once; we have $\Theta_\id = 1$. The $T$ matrix has a prefactor
$e^{-i \frac{2\pi}{24} c} = T_{\id \id}$, where $c$ is the central charge of the anyon model.
The $S$-matrix contains the mutual statistics, i.e. the statistics obtained when particles
of general types $a$, $b$ are braided around each other. The first row of the $S$
matrix contains the quantum dimensions $d_a$ of the quasiparticles.

To numerically obtain these matrices, first a quasi-orthonormal basis $\{ \ket{\Psi^\mathrm{tor}_a} \}$ on
a torus of $3 \times L \times L$ sites with well-defined topological flux $a$ through the torus and a
$\pi/3$ rotational symmetry must be obtained.
As pointed out in Ref.~\onlinecite{zhang2012}, the modular matrices $T$ and $S$ of the emergent anyon
model can then be extracted from the matrix elements of a $\pi/3$ rotation $R_{\pi/3}$ by using the relation
\begin{equation}
\langle \Psi^\mathrm{tor}_a | R_{\pi/3} | \Psi^\mathrm{tor}_b \rangle = (D T S^{-1} D^\dagger)_{ab}.
\end{equation}
Here, $D$ is a diagonal matrix containing only complex phases; it accounts for the freedom in choosing the phases of the vectors $|\Psi^\mathrm{tor}_a\rangle$.
In Ref.~\onlinecite{cincio2012}, it was shown
(i) how to build the basis $|\Psi^\text{tor}_a\rangle$ from an MPS representation for $|\Psi^\text{cyl}_a\rangle$ and
(ii) how to extract both $T$ and $S$ from $\langle \Psi^\mathrm{tor}_a |R_{\pi/3}| \Psi^\mathrm{tor}_b \rangle$.
Note that the matrix $T$ is referred to as $U$ in the reference.

%


\section{Method details}

\subhead{Finite-size DMRG}
To study the finite cylinders and strips discussed in the main text that are beyond the system size
amenable to exact diagonalization, we use the density-matrix renormalization group (DMRG) method
~\cite{white1992,white1992-1,schollwoeck2005,schollwoeck2011}.
While for many years limited to one-dimensional systems by its exponential scaling in the width of
quasi-one-dimensional systems, recent years have seen a surge in applications of large-scale
DMRG calculations to 2d systems~\cite{stoudenmire2012}.
A key point is that while it scales exponentially in the width of the system, the scaling is polynomial in the
length, allowing it to go much beyond exact diagonalization if boundary conditions and the mapping
to a one-dimensional system are chosen appropriately.
DMRG is based on a variational ansatz, matrix-product states~\cite{schollwoeck2011},
that can be systematically improved by increasing the number $M$ of states kept,
where the computational cost grows as $\mathcal{O}(M^3)$. For the calculations shown
in Fig.~\ref{fig:spectrum} of the main text, we use up to $M=3600$ states.

From DMRG, we can easily extract the entanglement entropy for a contiguous block of
sites at the end of an open system. This allows us to calculate the central
charge of a gapless quasi-one-dimensional system by performing a fit to the well-known
results of Refs.~\onlinecite{holzhey1994,calabrese2004}, which for the entanglement entropy
of a contiguous block of $n=N/2$ sites in an open $1d$ system of $N$ sites has
\begin{equation} \label{eqn:Sc}
S(N) = S_0 + \frac{c}{6} \log \left( \frac{2N} {\pi} \right).
\end{equation}

\subhead{Infinite-size DMRG}
As discussed in the Methods summary, we use infinite-size DMRG to efficiently obtain ground
states, parametrized through translationally invariant matrix-product states,
on infinite cylinders up to XC12-6 and with up to $M=4096$ states, exploiting a U(1)
symmetry of the system. The calculation of the entanglement spectrum as well as the modular
matrices has been explained in the Methods summary; below, we provide additional
detail on numerical parameter used. We also discuss how to extract the fidelity
and the bulk gaps from such an infinite ansatz.

\paragraph*{Modular matrices}
The approach to extracting the $T$ and $S$ matrix outlined in the Methods summary requires
the calculation of the matrix elements of $\langle \Psi^\mathrm{tor}_a |R_{\pi/3}| \Psi^\mathrm{tor}_b \rangle$,
where $R_{\pi/3}$ is a 60-degree rotation. The cost of an exact calculation of this matrix
scales as a very large power in the bond dimension $M$, and in practice we have to resort to sampling techniques
to obtain an approximate evaluation. The result of Eq.~\eqnref{numS} of the main text were obtained using $2 \cdot 10^5$ samples
over spin configurations in the $S_z$ basis using a Markov chain Monte Carlo update scheme.
We have checked that the values for $T$ and $S$ are converged to within $10^{-3}$.
Therefore, the main source of the deviation between the numerically obtained results and the exact matrices is due to finite size effects. 

\paragraph*{Fidelity}
While the calculation of the fidelity
\begin{equation}
F(\theta) = \langle \Psi_a(\theta-\epsilon) | \Psi_a(\theta+\epsilon) \rangle
\end{equation}
is straightforward in the case of finite-size DMRG and exact diagonalization, more care must be taken in the case of infinite-size DMRG as all states are either equal or orthogonal in the thermodynamic limit. In the case of translationally invariant infinite MPS, one can consider the spectrum of the transfer operator of the product of two normalized infinite MPS. Given the eigenvalue with largest absolute value $|\lambda| \in [0,1]$ of this transfer operator, the overlap of the states for $N$ sites is $|\lambda|^N$. Clearly, as $N \rightarrow \infty$, $\lambda^N \rightarrow 0$ or 1. Therefore, instead of the overlap we quote directly the absolute value of the eigenvalue $|\lambda|$, which shows the same behavior as the fidelity for finite systems. We use $\epsilon = 0.025$.

\paragraph*{Bulk gaps}
In standard DMRG calculations, the triplet gap is easily available by changing
the total spin quantum number of the target state, which is enforced exactly. The
singlet gap can be obtained by first converging the ground state of the system and
then optimizing a second state under the constraint of orthogonality against the
ground state. For infinite-size DMRG calculations, however, neither of these methods
are applicable, and we therefore must resort to a different approach. For this
manuscript, we choose to approximately extract the gaps from the spectrum of
an effective local Hamiltonian obtained by contracting a network of the MPS with
the matrix-product operator representation of the Hamiltonian everywhere except
on one bond.

\begin{figure}
  \includegraphics[width=\columnwidth]{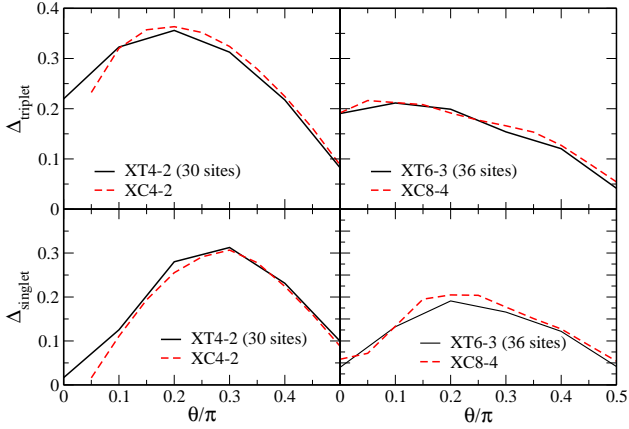}
  \caption{
  Comparison of singlet and triplet gaps obtained for infinite cylinders (XC8-4 and XC4-2) as well
  as for finite tori using DMRG and exact diagonalization of up to $N=36$ sites. All infinite-size results have been
  extrapolated to the $M \rightarrow \infty$ limit.
  \label{fig:gaps} }
\end{figure}

To validate our approach, we compare the energy gaps extracted from infinite cylinders
to small tori of the same diameter. Note that a comparison against finite-size DMRG on
long cylinders is hampered by the gapless edge that emerges in that topology.
In Fig.~\ref{fig:gaps}, we compare
(i) infinite XC4-2 cylinders against an XT4-2 torus with 30 sites, and
(ii) infinite XC8-4 cylinders against an XT6-3 torus with 36 sites.
The energies for the tori are mostly obtained with DMRG using up to $M=4000$ states; for both sizes,
we have performed exact diagonalization calculations at selected values of $\theta$ to confirm the ground state
energies obtained from DMRG are within 1\% error of the exact energies. The good agreement between
the two methods clearly validates the approach used.
We have also verified for various values of $\theta$ that for XC8-4, the gaps above the two ground states corresponding to
different topological sectors are comparable.
We therefore only show results for infinite cylinders in the main text.

\section{Perturbations}

\begin{figure}
  \includegraphics[width=\columnwidth]{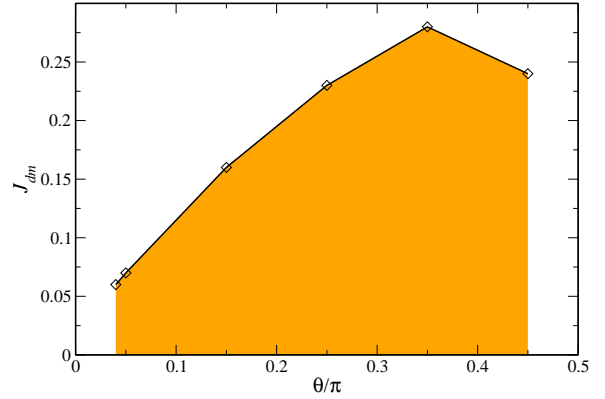}
  \caption{Phase diagram in the presence of DM interactions~\eqnref{eqn:dm}.
  \label{fig:dm} }
\end{figure}

Beyond the terms of Hamiltonian~\eqnref{eqn:ham} of the main text, we also consider two additional perturbations,
namely a particular form of the Dzyaloshinskii-Moriya (DM) interaction~\cite{moriya1960,dz1964,elhajal2002}
and a next-nearest neighbor (NNN) Heisenberg term. We explore these phases at a smaller bond dimension
of $M=256$ using infinite-size DMRG.

The DM interaction preserves lattice and U(1) spin symmetry, but breaks SU(2), and is given as
\begin{equation} \label{eqn:dm}
H_{DM} = J_{DM} \sum_{i < j} \hat{z} \cdot (S_i \times S_j).
\end{equation}
Here, the sum runs over nearest neighbors and clockwise around triangles. The phase diagram as a function
of DM interaction is shown in Fig.~\ref{fig:dm}. We observe that the phase is robust against
a DM term of strength comparable to the gap.

The simulation in the presence of the NNN Heisenberg term is much more computationally
challenging than the nearest-neighbor model considered above. We therefore have determined
the stability only for $\theta=0.15 \pi$. Using the fidelity as indicator of a phase transition, we have
found the phase to be stable for $J_\text{NNN} \in [-0.1,0.27]$.

We can break the SU(2) symmetry of the Heisenberg term without breaking lattice symmetries
by introducing an easy-axis anisotropy,
\begin{equation} \label{eqn:xxz}
H_\text{XXZ} = J_z \sum_{\langle i,j \rangle} S_i^z S_j^z.
\end{equation}
For negative values of $J_z$, we find that the phase is stable for all $J_z > -1.2$. The chiral phase appears
very stable for $J_z > 0$, and we have not been able to reliably determine a phase boundary.

\section{Network model}

\begin{figure}
  \includegraphics{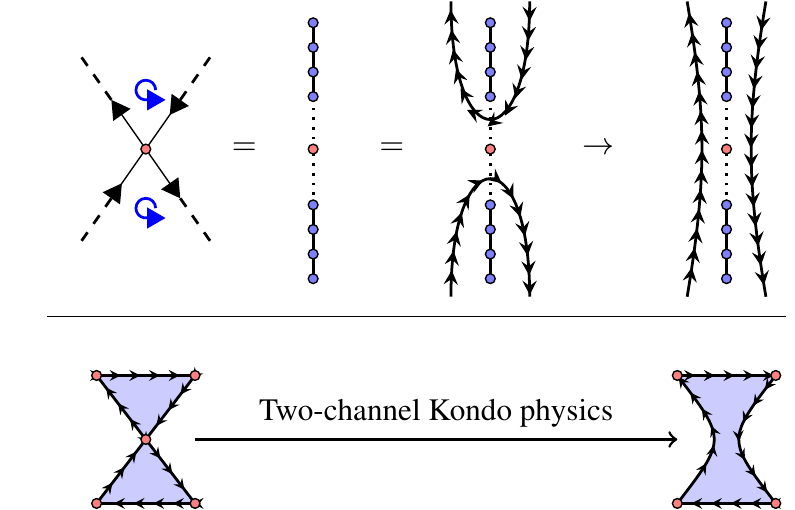}
  \caption{
  {\it Top panel:} Illustration of the behavior of the edge states at a corner shared by two puddles.
  {\it Bottom panel:} Behavior of two corner-sharing triangular puddles of the topological phase.
   \label{fig:puddles_supp} }
\end{figure}

We now revisit the network model perspective discussed in the main
text, and provide additional detail on how the {\it healing} of the edge states
surrounding two corner-sharing triangles that share one spin can
be understood in terms of two-channel Kondo physics.

If we envision the puddles to be very large, they would carry the edge
state on each side and the corner would look as shown in panel (I.a)
of Fig.~\ref{fig:puddles_supp}. The pair of edge states on the upper triangle is
known~\cite{halperin1983,moore1991} to be described by the same
theory as the right- and left-movers of a semi-infinite uniform spin-1/2
Heisenberg chain, and analogously for the lower pair of edge states.
The spin at the corner then appears as the center spin of an infinite
chain (panels (I.b,I.c)). It is well known that the infinite chain will heal
if the center spin is coupled to the two semi-infinite chains with equal
strength~\cite{eggert1992,kane1992}. Then, the right- and left-movers
will extend throughout the entire, infinite system (panel (I.d)).
The effect on the corner spin is summarized in panel II of Fig.~\ref{fig:puddles_supp},
where the situation shown in II.a corresponds to I.a, while II.d
corresponds to I.d. As is evident from II.d, the corner spin has
merged the two triangles to form a larger puddle encircled by
a single edge state, i.e. to form a larger region of the topological phase.

We can illustrate the validitiy of the above network model picture by
considering the situation where the spins are replaced by Majorana
fermion zero modes. Specifically,
we can form a term analogous to the spin chirality, a chiral hopping term
$\tilde{\chi}_{ijk} = \imath (\gamma_i \gamma_j + \gamma_j \gamma_k + \gamma_k \gamma_i)$.
This model, which is quadratic in fermionic operators, can be diagonalized straightforwardly.
Since the Kagome lattice is obtained as a triangular lattice of three-site
unit cells, we obtain three energy bands $E_\alpha(k_x,k_y)$, $\alpha = 1,2,3$.
We observe that all three bands are separated
by a gap from each other, and the central band $E_2$ is dispersionless,
i.e. $E_2(k_x,k_y) = 0$. Noting that the number of states in this band
coincides with the number of hexagons in the system, we identify these
zero-energy states as the non-interacting edge states encircling the
hexagons of the Kagome lattice shown in the right panel of Fig.~\ref{fig:kagome}.
We calculate the Chern number $C$~\cite{thouless1982}
using the real-space method of Refs.~\onlinecite{hastings2010,loring2010}
and find that $C$ of the top and bottom band is
$C=\pm 1$, i.e. that the model is in a topological phase; this was
previously observed in Ref.~\onlinecite{ohgushi2000}.

\bibliography{csl}

\end{document}